\documentclass[preprint,aps,pra,showpacs,floatfix]{revtex4}
\usepackage[dvips]{graphicx}
\usepackage{longtable}
\usepackage{dcolumn}
\usepackage[dvips]{graphicx}
\usepackage{bm}
\usepackage{bbm}
\usepackage{times}
\usepackage{nicefrac}
\usepackage{amsmath}
\usepackage{amsfonts}
\usepackage{amssymb}
\usepackage{amsthm}
\newcolumntype{.}{D{x}{}{-1}}

\begin{document}
\newcommand{\la}{\langle}
\newcommand{\ra}{\rangle}
\newcommand{\beq}{\begin{equation}}
\newcommand{\eeq }{\end{equation}}
\newcommand{\beqn}{\begin{eqnarray}}
\newcommand{\eeqn }{\end{eqnarray}}
\title{QED calculation of the $2p_{3/2}-2p_{1/2}$ 
transition energy in five-electron ion of argon}
\author{A.~N.~Artemyev,$^{1,2}$ V.~M.~Shabaev,$^{1,2}$ I.~I.~Tupitsyn,$^{1,2}$
G. Plunien,$^{2}$ and V.~A.~Yerokhin $^3$}
\affiliation{$^{1}$Department of Physics, St. Petersburg State University,
Oulianovskaya 1, Petrodvorets, St. Petersburg 198504, Russia\\
$^2$Institut f\"{u}r Theoretische Physik, TU Dresden, Mommsenstra{\ss}e 13,
D-01062 Dresden, Germany\\
$^3$ Center for Advanced Studies, St. Petersburg State Polytechnical
University, Polytekhnicheskaya 29, St. Petersburg 195251, Russia
}
\begin{abstract}
We perform {\it ab initio} QED calculation of the
$(1s)^2(2s)^22p_{3/2}$ - $(1s)^2(2s)^22p_{1/2}$ transition energy in
the five-electron ion of argon. The calculation is carried out by 
perturbation theory starting with an effective screening potential 
approximation. Four 
different types of the screening potentials are considered. The rigorous
QED calculations of the two lowest-order QED and electron-correlation
effects are combined with  approximate evaluations of the third- and 
higher-order electron-correlation contributions. The theoretical value for
the wavelength obtained amounts to 441.261(70) (nm, air) 
and perfectly agrees with the experimental one, 441.2559(1) (nm, air).
\end{abstract}

\maketitle
In the recent paper \cite{draganic:03}, high precision measurements of
the $2p_{3/2}-2p_{1/2}$ transition energy in Ar XV, Ar XIV, Ar XI, and
Ar X have been reported. The best results have been achieved for
the five-electron ion of argon, where the experimental precision is 2700 times
better than the theoretical one and 200 times
better than that in the best of the previous measurements. This 
unprecedent precision provides a unique possibility to test various branches 
of the theory describing many-electron systems. 

Since within 
the framework of the non-relativistic theory the $2p_{3/2}$ and $2p_{1/2}$
levels are degenerate, the transition energy is exclusively determined by the 
relativistic and quantum electrodynamic (QED) effects. Hence, 
investigation of this transition allows us to test 
the many-electron QED effects as well as  
calculations of  the relativistic electron-correlation effects up to an 
extremely high level of accuracy. 

To date {\it ab initio} calculations of many-electron QED effects were
considered for two- and three-electron ions 
\cite{artemyev:05, yerokhin:01:pra2p}. For systems with a larger number of
electrons these effects were accounted for only within some one-electron or
semi-empirical approximations 
\cite{safronova:96, draganic:03, indelicato:05}. 
Although the agreement of the results of Refs. 
\cite{draganic:03, indelicato:05} with experiment 
is rather good, the precision of these calculations
is much worse compared to the experimental one. 
The improvement of this precision is a challenging problem for 
the theory. The goal of the present letter is to improve the theoretical 
precision for the $2p_{3/2}-2p_{1/2}$ transition energy in Ar$^{13+}$.
To achieve this goal we perform 
rigorous QED calculations to first two orders
of perturbation theory and extremely large-scale configuration-interaction 
Dirac-Fock-Sturm (CI-DFS) calculations of the third- and higher-order 
contributions within the Breit approximation. 

To formulate the QED perturbation theory we use the two-time Green
function method \cite{shabaev:02:rep}.
Instead of usual Furry picture, where only the nucleus is 
considered as a source of the external field, we have used  an 
extended version of the Furry picture. It implies incorporation 
of some screening 
potential, which partly accounts for the
interelectronic interaction, into the zeroth-order
Hamiltonian. The perturbation theory is
formulated in powers of the difference between the full QED
interaction Hamiltonian and the screening potential. This accelerates the
convergence of the perturbative series. In addition, the usage of the screening
potential allows us to avoid the degeneracy of the $(1s)^2(2s)^22p_{3/2}$ and 
$(1s)^2(2p_{1/2})^22p_{3/2}$ states that occurs if the pure Coulomb potential 
is employed as the zeroth-order approximation. 

In the present letter we use four different types of the effective potential. 
The simplest choice is the core-Hartree (CH) potential. 
To obtain this potential we add  
the radial charge density distribution of the 
(four) core electrons 
\beq
\rho_{\rm c}=2\sum_{n=1s,2s}
(g_n^2+f_n^2) 
\eeq
with $g$ and $f$ being the upper and lower radial components of the one-electron 
Dirac wave function 
to the radial charge density distribution $\rho_{\rm nuc}$ of the nucleus.  
The nuclear charge density is described by a Fermi distribution. 
The potential $V_{\rm CH}$ generated by
the total charge density $\rho=\rho_{\rm nuc}+\rho_{\rm c}$ is calculated 
self-consistently by solving the Dirac equation.

The second choice is a local potential derived by inversion of the radial 
Dirac equation with the wave function obtained by solving the Dirac-Fock (DF)
equation \cite{shabaev:05:pnc}. 
We will refer to this potential as local Dirac-Fock (LDF) 
potential.
The construction of the potential $V_{\rm LDF}$ is described in details in
Ref. \cite{shabaev:05:pnc}.

The other two potentials are based on the results of the 
density-functional theory
(DFT). The first one is referred to as the
Slater potential \cite{slater:51}. This potential belongs to the wide family of
$x_\alpha$ potentials.
Introducing the total one-electron radial density $\rho_{\rm t}$ via 
\beqn
\rho_{\rm t}(r)&=&4\pi r^2 \rho(r)\,,\\
\int \rho(r) d^3r&=&\int_0^\infty \rho_{\rm t}(r)dr=N\,,
\eeqn
where $N$ is the total number of the electrons, one can write the 
$x_\alpha$ potential in simple form:
\beq\label{xalpha_potl}
V(r)=V_{\rm nuc}(r)+\alpha\int_0^\infty \frac{\rho_{\rm t}(r)}{r_>}-
x_\alpha\frac \alpha r \left(\frac{81}{32\pi^2}r\rho_{\rm t}(r)\right)^{1/3}\,.
\eeq
Here $\rho_{\rm t}$ denotes the total one-electron density, 
i.e. includes both core-electron and valence-electron density, 
while the CH potential includes only the core-electron density.
The value of the constant
$x_\alpha$ for the case of the Slater potential is equal to $1$.
To improve the asymptotic behavior of this potential at large distances,
a self-interaction correction, known also as the Latter  
correction \cite{latter:55}, has been added to it. 

The fourth potential used in our calculations 
is known as Perdew-Zunger potential $V_{\rm PZ}$.
It was constructed as described in Ref. \cite{perdew:81}. 

The calculations of the transition energy can be conveniently
divided in several steps. At first one has to solve the 
Dirac equation with the effective potential. 
Bound-state QED calculations require the representation of 
the quasi-complete set of the Dirac equation solutions. 
This was achieved by employing the dual-kinetic-balance (DKB) finite basis set 
method \cite{shabaev:04:dkb} with basis functions constructed from B-splines
\cite{johnson:88}.

Next we calculate the set of Feynman diagrams shown in Fig. 1 
without any photon or electron
loop, i.e. the part describing the interelectronic interaction. 
The dashed line ending with a triangle represents the interaction with the 
screening potential, taken with the opposite sign. 
The formulas for the calculations of the diagrams (a)-(d) in the 
framework of QED can be found in our previous works 
(see, e.g., Ref. \cite{yerokhin:01:pra2p}) devoted to the 
calculations of the two-photon exchange corrections to the energy levels 
of Li-like ions. We note that the diagrams containing only core electrons as initial 
(final) states can be omitted, because their contributions do not affect the
transition energy. The formulas from Ref. \cite{yerokhin:01:pra2p} in
our case should be completed by similar ones with the $1s$ state
being replaced by the $2s$ state. 
For the contributions of the diagrams (e)-(g) one can obtain:
\beqn
\Delta E_e&=&V_{vv} \,,\\
\Delta E_f&=&\sum_{n\ne v} \frac{V_{vn}^2
}{\varepsilon_v-\varepsilon_n}\,, \\
\Delta E_g&=&2\sum_{c=1s,2s;\mu_c}\left[\sum_{n\ne v} \sum_P (-1)^P
\frac{I_{PcPvcn}(\varepsilon_{Pc}-\varepsilon_c)V_{nv}}
{\varepsilon_v-\varepsilon_n} \right .\nonumber \\ &&+\left .
\sum_{n\ne c} \sum_P (-1)^P
\frac{I_{PcPvnv}(\varepsilon_{Pv}-\varepsilon_v)V_{nc}}
{\varepsilon_c-\varepsilon_n}\right]\nonumber \\
&&-\sum_{n=c} (V_{vv}-V_{nn})I'_{vnnv}(\varepsilon_v-\varepsilon_n)
\,,
\eeqn
where $V_{ab}=-\la
a | V_{\rm scr}|b\ra$, $I_{abcd}(\omega)=\la ab|I(\omega)|cd\ra$, 
$I(\omega)=e^2\alpha^\mu \alpha^\nu D_{\mu\nu}(\omega)$, $D$ is the
photon propagator, $P$ is the permutation operator, $(-1)^P$ is the
sign of the permutation, 
$I'_{abcd}(\omega)=\la ab|\frac{\partial}{\partial \omega}I(\omega)|cd\ra$, 
and the indices $c$ and $v$ denote the wave
functions of the core ($1s$ or $2s$) and the valence state,
respectively.
 
In the next step we should take into account the contribution of the
diagrams depicted in Fig. \ref{fig:QED}. 
The evaluation of the one-electron QED corrections of order $\alpha$, i.e.  
the  self energy (SE) and the vacuum polarization (VP) (part (a) and (f) in Fig.  \ref{fig:QED}) 
in an external Coulomb field is well known 
\cite{mohr:74, soff:88:vp, manakov:89:zhetp, mps:98}. 
High precision calculation of these diagrams for
the case of an arbitrary external field, however, is a more involved
problem \cite{cheng:93, lindgren:93:pra}. We adopted the finite basis
method for such calculations. 

The main QED contribution to the energy of the forbidden transition arises
from the one-electron SE diagram. To reach the required precision, this
contribution has to be evaluated with at least 0.1\% of accuracy. 
With this purpose, we decompose the SE diagram into
zero-, one-, two-, and many-potential terms, as 
indicated in Fig. \ref{fig:eqn}, where
the dashed line ended with a 
cross denotes the full effective external potential. The zero- and 
one-potential terms have been calculated in momentum-space representation using the
traditional renormalization scheme. The two-potential term has been evaluated
in the coordinate space employing the analytical representation of the 
partial-wave decomposition of the free-electron Green function. 
To reach the required accuracy necessitates to sum over partial waves up to 
angular quantum number $\kappa = 50$. 
Corresponding calculation within the B-spline approach
would require an enormously large number of the basis functions, which would 
slow-down the computation considerably. 
To circumvent this problem 
we extracted the slowly-converging two-potential term and calculated 
it separately. 
A new numerical technique was elaborated for performing 
this calculation and will be described in details in our following work. 
The remaining many-potential 
term, containing three and more potentials, has been calculated within the
DKB approach as the difference between the term, containing two- and more
potentials and the two-potential one. 
The convergence of the partial-wave
series for this term is very fast and, to reach the required accuracy, 
the summation can be restricted to $|\kappa| \le 10$.

Since the contribution of the diagrams involving VP loops (parts (f--j) of
Fig.  \ref{fig:QED})
was found to be very small, we calculated it within the Uehling
approximation. The total VP contribution to the
transition energy in the Uehling approximation is 
about 0.2 cm$^{-1}$. The remaining Wichmann-Kroll contribution is 
negligible.

The other diagrams in Fig. \ref{fig:QED} (b--e) 
represent the
self-energy screening contributions. The irreducible part of the
diagrams (d) and (e) can be calculated as a wave function correction to the
one-electron SE diagram. This part was calculated using the same algorithm as
it was used for the calculations of the one-electron SE diagram. 
The diagrams (b) and
(c) are known as vertex diagrams. They were calculated in the traditional way: The
bound-electron propagator was decomposed into zero- and many-potential
terms. The zero-potential term is ultraviolet divergent. It 
was renormalized and calculated in  momentum space 
 together with the zero-potential term of the reducible part of the
diagrams (d) and (e). 
The remaining many-potential term contains infrared divergent terms. These
divergences are canceled by the infrared divergences in the reducible part of
the many-potential term of the diagrams (d) and (e) (see, e.g., Ref. 
\cite{yerokhin:99:scr}).
All the many-potential terms were calculated using the DKB approach. 

\begin{figure}[ht]
\centering
\includegraphics[clip=true,width=0.7\columnwidth]{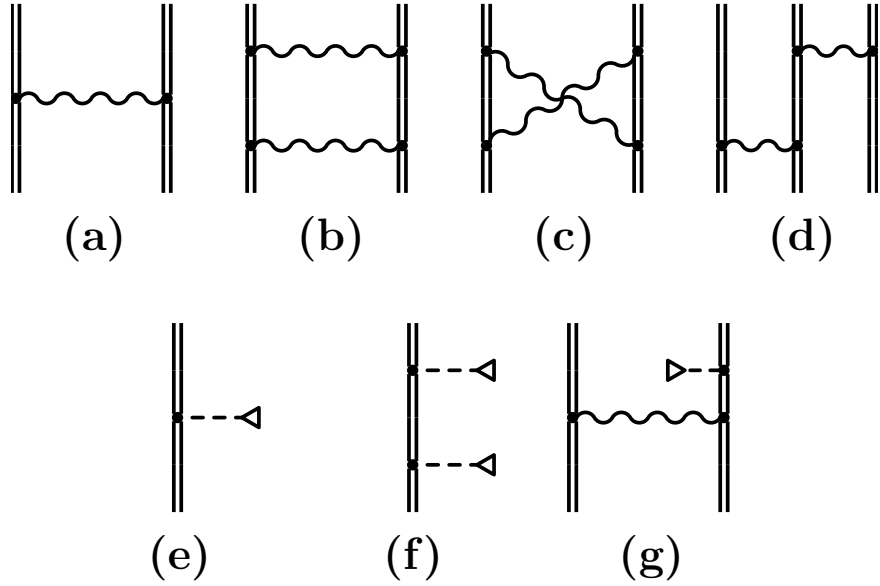}
\caption{Interelectronic-interaction diagrams. 
\label{fig:interaction}}
\end{figure}

\begin{figure}[ht]
\centering
\includegraphics[clip=true,width=0.7\columnwidth]{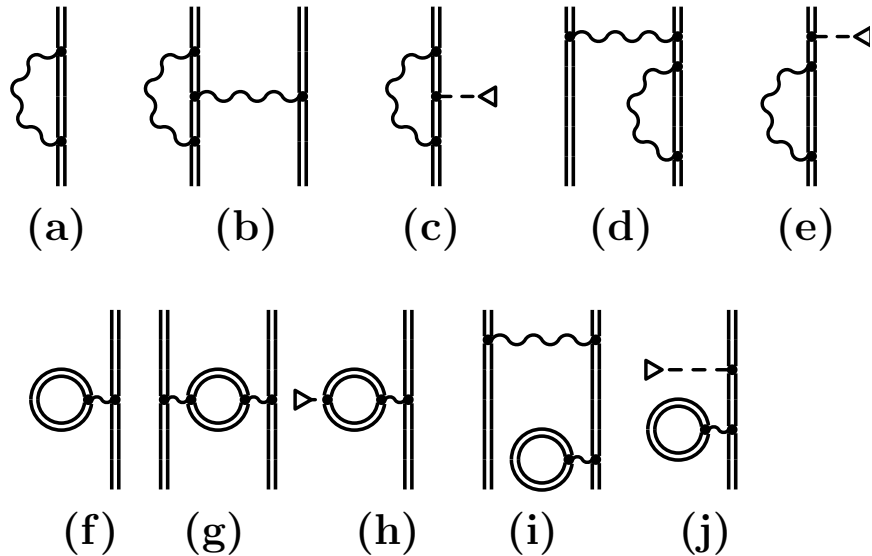}
\caption{Self-energy and vacuum-polarization diagrams. 
\label{fig:QED}}
\end{figure}

\begin{figure}[ht]
\centering
\includegraphics[clip=true,width=0.7\columnwidth]{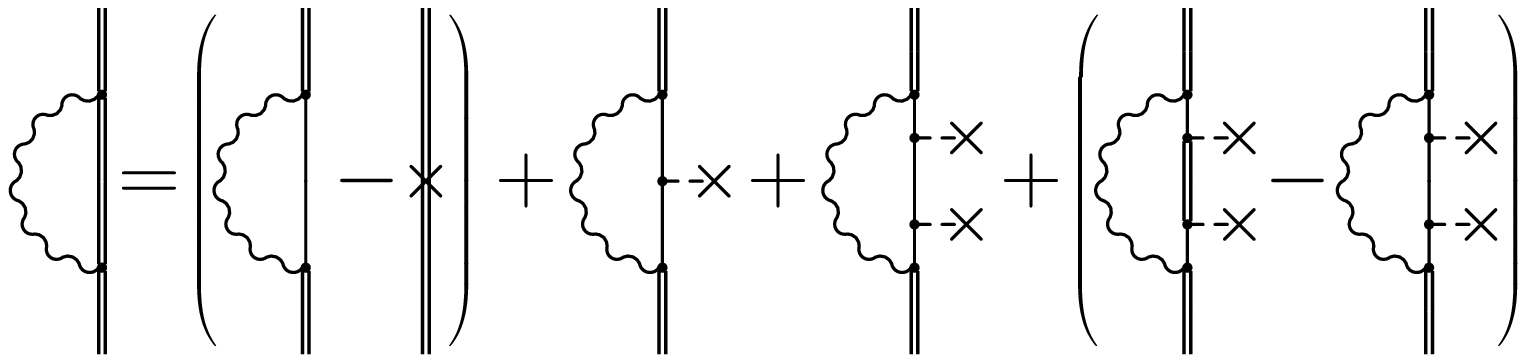}
\caption{Diagram equation for the calculation of the one-electron
  self-energy contribution. 
\label{fig:eqn}}
\end{figure}

The contribution of the interelectronic-interaction diagrams 
of the third and higher orders has been
evaluated within the Breit approximation using the CI-DFS method. Going beyond 
the calculation performed in Ref. \cite{draganic:03}, the basis 
set of the configuration state functions
was significantly enlarged and the quadruple excitations were included. 
To separate out  the contribution of the third- and higher-order 
diagrams from the total energy, 
the following algorithm has been used. First of 
all, the Hamiltonian of the CI-DFS calculations was decomposed into parts 
\beqn
H&=&H_0+\lambda V\, ,\\
H_0&=&H_{\rm free}+V_{\rm nuc}+V_{\rm scr}\, ,\\
V&=&V_{ee}-V_{\rm scr}\,,
\eeqn
where $H_0$ is the unperturbed (zero-order) Hamiltonian, 
$V_{\rm nuc}$ and $V_{ee}$ denote the operators of the electron-nucleus and 
electron-electron (Coulomb and Breit)
interaction, respectively. $V$ defines a perturbation 
 and $\lambda$ is a freely varying parameter. 
This representation allows us to perform the expansion of the energy $E$ in
powers of $\lambda$
\beq \label{expansion}
E(\lambda )=E^{(0)}+\lambda E^{(1)}+\lambda^2 E^{(2)}+ \lambda^3 E^{(\ge 3)}(\lambda)\,.
\eeq
It is easy to see that the coefficients $E^{(1)}$ and $E^{(2)}$ correspond 
to the first- and second-order diagrams depicted in Fig. 
\ref{fig:interaction}, calculated within the Breit approximation.
Utilizing the same basis set 
these coefficients have been evaluated in two
different ways: via numerical differentiation of
$E(\lambda)$ with respect to $\lambda$ evaluated at $\lambda =0$ and directly by
means of perturbation theory. 
The term $E^{(\ge 3)}$ is then calculated  
as the difference between the total energy and the first three terms in 
Eq. (\ref{expansion}) evaluated at $\lambda = 1$. 
The uncertainty of the $E^{(\ge 3)}$ contribution obtained in this way is 
estimated to be about 3 cm$^{-1}$. The QED contributions of the third and 
higher orders, which are beyond the Breit approximation, have not yet been 
evaluated. We estimate the uncertainty due to these effects as $\pm 2$ 
cm$^{-1}$.

The contribution of
the one-electron two-loop QED-radiative corrections 
is very small and can be estimated 
using the analytical $\alpha Z$-expansion reported in  
Ref. \cite{jentschura:05}. 
Finally, one has to take into account
the nuclear recoil effect. The
calculation of this effect to all orders in
$\alpha Z$ was performed in our recent paper
\cite{orts:06}, where the isotope shift of the forbidden
transition energies in B- and Be-like argon has been investigated. 

The results of our calculations are presented in Table 1. 
In the first line of the table the difference between the one-electron Dirac
energies of $2p_{3/2}$ and $2p_{1/2}$ states,
calculated for the different screening potentials,
is given. In the second row we
give the contribution of first-order diagrams from  
Fig. \ref{fig:interaction} (diagrams (a) and (e)). 
These diagrams are calculated in
the rigorous framework of QED, i.e. taking into account the energy 
dependence of the photon propagator. In the third line the contribution 
of the diagrams of the second order from Fig. \ref{fig:interaction} calculated
within the Breit approximation is given. 
In the fourth line we give the QED correction to this contribution, i.e. the
difference between these diagrams evaluated within the 
rigorous QED approach and within the 
Breit approximation. In the fifth line we give the contribution of the
interelectronic-interaction diagrams of the third and higher orders derived
from the CI-DFS calculations as described above. 
The contribution of the first- and second-order diagrams from 
Fig. \ref{fig:QED} are presented in the sixth and seventh lines, respectively.
The uncertainty due to uncalculated third- and higher-order QED effects is 
presented in eighth line. 
The following lines compile the contribution of the
two-loop one-electron QED and the nuclear recoil correction, respectively.
Finally, in the last line we present the total
values of the energy of the forbidden $2p_{3/2}-2p_{1/2}$ transition in B-like
argon, calculated for the four different screening potentials. 
Averaging these values and accounting for the uncertainty due to
the higher-order interelectronic-interaction and QED effects, we obtain  
$E_{\rm tot}=$22656.1(3.6)cm$^{-1}$. This value is four times more
precise than that of Ref. \cite{draganic:03}:
22662(14)cm$^{-1}$. The improvement mainly results from the calculation of 
the second-order QED effects (lines 4 and 7 in the table) as well as from
the much more elaborated CI-DFS calculations performed in this work.
Converting our result to the wavelength in air 
with the aid of Ref. \cite{stone_inet},
one can obtain 441.261(70) (nm, air), 
which perfectly agrees with the experimental 
result 441.2559(1) (nm, air) from Ref. \cite{draganic:03}.
Further improvement of the theoretical value can be achieved by the
reducing of the uncertainty of the CI-DFS calculations and computation of the
QED diagrams of the third and higher orders.

Summarizing, in this work we have calculated the energy of 
the forbidden $2p_{3/2}-2p_{1/2}$ transition in the five-electron ion of
argon. The calculation incorporates the rigorous treatment of the 
second-order many-electron QED effects and the large-scale CI-DFS calculations 
of the third- and higher-order electron-correlation effects.
This is the first {\it ab initio} treatment of the five-electron system in
framework of QED. Significant improvement of the agreement between 
theoretical and experimental data has been achieved.

This work was partly supported by RFBR grant No. 04-02-17574. 
A.N.A. acknowledges the support by INTAS YS grant No. 03-55-960 and by the 
"Dynasty" foundation. V.M.S., I.I.T., and G.P. acknowledge the support by 
INTAS-GSI grant No. 06-1000012-8881.
G.P. also acknowledges support from DFG and GSI. 
\begingroup
\begin{table}[ht]
\caption{Various contributions to the energy of the forbidden transition in
  B-like argon and the total result (in cm$^{-1}$).
 \label{tab:QED}}
\begin{ruledtabular}
\begin{tabular}{l....}
   & V_{\rm CH} & V_{\rm LDF}&V_{\rm PZ}&V_{\rm Sl} \\
\hline
$E_{\rm Dirac}$ &
24343.x0&25276.x8&24860.x7&26482.x6\\
$E^{(1)}_{\rm int}$ &
-1002.x9&-2064.x2&-1539.x3&-3451.x8\\
$E^{(2)}_{\rm int, Breit}$ &
-1428.x4&-1215.x4&-1091.x8&-993.x3\\
$E^{(2)}_{\rm int, QED}$&
6.x2&5.x5&5.0&8.x7\\
$E^{(\ge 3)}_{\rm int, Breit}$&
694.x0(3.0)&608.x2(3.0)&376.x0(3.0)&567.x6(3.0)\\
$E^{(1)}_{\rm QED}$ &
47.x9&49.x9&48.x9&52.x5\\
$E^{(2)}_{\rm QED}$ &
-2.x2&-5.x4&-3.x8&-8.x2\\
$E^{(\ge 3)}_{\rm QED}$ &
\pm 2.x0&\pm 2.x0&\pm 2.x0&\pm 2.x0\\
$E_{\rm two-loop}$&
-0.x1(1)&-0.x1(1)&-0.x1(1)&-0.x1(1)\\
$E_{\rm recoil}$&
-0.x6&-0.x6&-0.x6&-0.x6\\
$E_{\rm total}$ &
22656.x9&22654.x8&22655.x2&22657.x4\\
\end{tabular}
\end{ruledtabular}
\end{table}
\endgroup

\begin{thebibliography}{10}
\bibitem{draganic:03}
I.~Dragani\'{c}, J.~R.~Crespo L\'{o}pez-Urrutia, R.~DuBois, S.~Fritzsche, V.~M.
  Shabaev, R.~Soria Orts, I.~I. Tupitsyn, Y.~Zou, and J.~Ullrich,
\newblock Phys. Rev. Lett. {\bf 91}, 183001 (2003).
\bibitem{artemyev:05}
A.~N. Artemyev, V.~M. Shabaev, V.~A. Yerokhin, G.~Plunien, and G.~Soff,
\newblock Phys. Rev. A {\bf 71}, 062104 (2005).
\bibitem{yerokhin:01:pra2p}
V.~A. Yerokhin, A.~N. Artemyev, V.~M. Shabaev, M.~M. Sysak, O.~M. Zherebtsov,
  and G.~Soff,
\newblock Phys. Rev. A {\bf 64}, 032109 (2001).
\bibitem{indelicato:05}
P.~Indelicato, E.~Lindroth, and J.~P. Desclaux,
\newblock Phys. Rev. Lett. {\bf 94}, 013002 (2005).
\bibitem{safronova:96}
M.~S. Safronova, W.~R. Johnson, and U.~I. Safronova,
\newblock Phys. Rev. A {\bf 54}, 2850 (1996).
\bibitem{shabaev:02:rep}
V.~M. Shabaev,
\newblock Physics Reports {\bf 356}, 119  (2002).
\bibitem{shabaev:05:pnc}
V.~M. Shabaev, I.~I. Tupitsyn, K.~Pachucki, G.~Plunien, and V.~A. Yerokhin,
\newblock Phys. Rev. A {\bf 72}, 062105 (2005).
\bibitem{slater:51}
J.~C. Slater,
\newblock Phys. Rev. {\bf 81}, 385  (1951).
\bibitem{latter:55}
R.~Latter,
\newblock Phys. Rev. {\bf 99}, 510  (1955).
\bibitem{perdew:81}
J.~P. Perdew and A.~Zunger,
\newblock Phys. Rev. B {\bf 23}, 5048  (1981).
\bibitem{shabaev:04:dkb}
V.~M. Shabaev, I.~I. Tupitsyn, V.~A. Yerokhin, G.~Plunien, and G.~Soff,
\newblock Phys. Rev. Lett. {\bf 93}, 130405 (2004).
\bibitem{johnson:88}
W.~R. Johnson, S.~A. Blundell, and J.~Sapirstein,
\newblock Phys. Rev. A {\bf 37}, 307  (1988).
\bibitem{mohr:74}
P.~J. Mohr,
\newblock Ann. Phys. (New York) {\bf 88}, 26, 52  (1974).
\bibitem{soff:88:vp}
G.~Soff and P.~J.~Mohr,
\newblock Phys. Rev. A {\bf 38}, 5066 (1988).
\bibitem{manakov:89:zhetp}
N.~L. Manakov, A.~A. Nekipelov, and A.~G. Fainshtein,
\newblock Zh. Eksp. Teor. Fiz. {\bf 95}, 1167 (1989),
\newblock [Sov. Phys. JETP {\bf 68}, 673 (1989)].
\bibitem{mps:98}
P.~J. Mohr, G.~Plunien, and G.~Soff,
\newblock Phys. Rep. {\bf 293}, 227 (1998).
\bibitem{cheng:93}
K.~T. Cheng, W.~R. Johnson, and J.~Sapirstein,
\newblock Phys. Rev. A {\bf 47}, 1817  (1993).
\bibitem{lindgren:93:pra}
I.~Lindgren, H.~Persson, S.~Salomonson, and A.~Ynnerman,
\newblock Phys. Rev. A {\bf 47}, R4555  (1993).
\bibitem{yerokhin:99:scr}
V.~A. Yerokhin, A.~N. Artemyev, T.~Beier, G.~Plunien, V.~M. Shabaev, and
  G.~Soff,
\newblock Phys. Rev. A {\bf 60}, 3522  (1999).
\bibitem{jentschura:05}
U.~D. Jentschura, A.~Czarnecki, and K.~Pachucki,
\newblock Phys. Rev. A {\bf 72}, 062102  (2005).
\bibitem{orts:06}
R.~Soria Orts, Z.~Harman, J.~R.~Crespo L\'{o}pez-Urrutia, A.~N. Artemyev, H.~Bruhns,
  A.~J.~G. Mart\'{i}nez, U.~D. Jentschura, C.~H. Keitel, A.~Lapierre,
  V.~Mironov, V.~M. Shabaev, H.~Tawara, I.~I. Tupitsyn, J.~Ullrich, and A.~V.
  Volotka,
\newblock Phys. Rev. Lett. {\bf 97}, 103002 (2006).
\bibitem{stone_inet}
J.~A. Stone and J.~H. Zimmerman,
\newblock http://emtoolbox.nist.gov .
\end{thebibliography}
\end{document}